%% file: oopsla-paper.tex
\documentclass[acmsmall]{acmart}
\acmJournal{PACMPL}
\acmVolume{1}
\acmNumber{CONF} % CONF = POPL or ICFP or OOPSLA
\acmArticle{1}
\acmYear{2018}
\acmMonth{1}
\acmDOI{} % \acmDOI{10.1145/nnnnnnn.nnnnnnn}
\startPage{1}

\setcopyright{none}
\usepackage{booktabs}   %% For formal tables:
\usepackage{subcaption} %% For complex figures with subfigures/subcaptions
\usepackage{makecell}
\usepackage{algorithmic}
\usepackage{graphicx}
\usepackage{pifont}
\usepackage{colortbl}
\usepackage{xspace}
\usepackage{wasysym}
\usepackage{multirow}
\usepackage[frozencache,cachedir=.]{minted}
\usepackage{todonotes}
\usepackage{stfloats}

\begin{document}

\newcommand{\tool}{PROMPT\xspace}
\newcommand{\Times}{\(\times\)\xspace}

\definecolor{ForestGreen}{RGB}{34,139,34}
\definecolor{YellowGreen}{rgb}{0.6, 0.8, 0.2}
\definecolor{BurntOrange}{rgb}{0.8, 0.33, 0.0}
\definecolor{OrangeRed}{rgb}{1.0, 0.27, 0.0}
  
\def\naive{na\"{\i}ve}
\def\Naive{Na\"{\i}ve}
\def\naively{na\"{\i}vely}
\def\Naively{Na\"{\i}vely}
\title{PROMPT: A Fast and Extensible Memory Profiling Framework}
\author{Ziyang Xu}
\affiliation{Princeton University\country{USA}}

\author{Yebin Chon}
\affiliation{Princeton University\country{USA}}

\author{Yian Su}
\affiliation{Northwestern University\country{USA}}

\author{Zujun Tan}
\affiliation{Princeton University\country{USA}}
\author{Sotiris Apostolakis}
\affiliation{Princeton University\country{USA}}
\author{Simone Campanoni}
\affiliation{Northwestern University\country{USA}}

\author{David I. August}
\affiliation{Princeton University\country{USA}}
\input{sections/abstract}

%% 2012 ACM Computing Classification System (CSS) concepts
%% Generate at 'http://dl.acm.org/ccs/ccs.cfm'.

%% End of generated code

%% Keywords
%% comma separated list
% \keywords{Memory Profiling, , keyword3}  %% \keywords are mandatory in final camera-ready submission

%% \maketitle
%% Note: \maketitle command must come after title commands, author
%% commands, abstract environment, Computing Classification System
%% environment and commands, and keywords command.
\maketitle

\input{sections/intro} 
\input{sections/background}
\input{sections/overview}
\input{sections/design}
\input{sections/implementation}
\input{sections/evaluation}
\input{sections/discussions}
\input{sections/related}
\input{sections/conclusion}

% %-------------------------------------------------------------------------------
% \section*{Acknowledgments}
% %-------------------------------------------------------------------------------

% The USENIX latex style is old and very tired, which is why
% there's no \textbackslash{}acks command for you to use when
% acknowledging. Sorry.

% %-------------------------------------------------------------------------------
% \section*{Availability}
% %-------------------------------------------------------------------------------

% USENIX program committees give extra points to submissions that are
% backed by artifacts that are publicly available. If you made your code
% or data available, it's worth mentioning this fact in a dedicated
% section.

% \bibliographystyle{ACM-Reference-Format}
\bibliography{reference}

%% Bibliography
%\bibliography{bibfile}

\end{document}

%% file: sections/abstract.tex
\begin{abstract}
Memory profiling captures programs' dynamic memory behavior, assisting
programmers in debugging, tuning, and enabling advanced compiler
optimizations like speculation-based automatic parallelization.
As each use case demands its unique program trace summary, various memory
profiler types have been developed.
Yet, designing practical memory profilers often requires extensive compiler
expertise, adeptness in program optimization, and significant implementation
efforts. This often results in a void where aspirations for fast and robust
profilers remain unfulfilled.
To bridge this gap, this paper presents PROMPT, a pioneering framework for
\textit{streamlined} development of \textit{fast} memory profilers.
With it, developers only need to specify profiling events and define the core
profiling logic, bypassing the complexities of custom instrumentation and 
intricate memory profiling components and optimizations.
Two state-of-the-art memory profilers were ported with
PROMPT while all features preserved. By focusing on the core profiling logic,
the code was reduced by more than 65\% and the profiling speed was improved by
5.3$\times$ and 7.1$\times$ respectively.  To further
underscore PROMPT's impact, a tailored memory profiling workflow was constructed
for a sophisticated compiler optimization client.  In just 570 lines of code,
this redesigned workflow satisfies the client's memory profiling needs while
achieving more than 90\% reduction in profiling time and improved robustness
compared to the original profilers.
\end{abstract}

%% file: sections/intro.tex
\section{Introduction}
\label{sec:intro}

Profiling techniques summarize runtime information of a specific run of the
program.  Examples of profile information include a summary of the hot regions
of the program, edge weights on the control flow, and the frequency of
manifested memory dependences.  Programmers use profiles to guide debugging and
tuning of programs.  Compilers use profiles to guide sophisticated program
optimizations~\cite{dehao:16:cgo, pogo:web, intelc++:web, g++:prof,
panchenko:2019:bolt}.  Memory profiling is a type of profiling concerned with
memory-related program behavior and is particularly useful in overcoming the
limitations of compiler memory analysis to unlock \textit{speculative}
transformations that can dramatically improve program
performance~\cite{connors:97:uiucece,
liu:06:ppopp, steffan:00:isca, bridges:07:micro,
thies:07:micro, wu:08:lcpc,johnson:12:pldi}. 
These speculative transformations optimistically assume memory behaviors
observed in profiling runs scale to production workloads.
They can also preserve correctness at runtime by executing recovery code when a
specific dynamic instance of an assumption is detected to be false.   Since
trends in profiled behavior tend to hold regardless of program input, the cost
of recovery code is low compared to the gains obtained by the unlocked
transformations.  Performance is shown to improve by orders of magnitude for
many programs~\cite{liu:06:ppopp, steffan:00:isca, wu:08:lcpc,
bridges:07:micro}.

Many types of memory profiling have been proposed to address various needs,
including memory dependence
profiling~\cite{larus:1993:pp,chen:04:cc,zhang:09:cgo}, value pattern
profiling~\cite{gabbay:1997:valuepred}, object lifetime
profiling~\cite{wu:2004:obj}, and points-to
profiling~\cite{johnson:2012:Privateer}. 
A memory profiler first tracks program events related to program's memory
behavior, like memory accesses, loop invocations, and function calls.  Then, it
uses the events to summarize the memory behavior in some way.  Both tracking and
summarizing are usually expensive. Thus, memory profilers must be heavily
optimized to be practical.  As a result, memory profiler developers must master
a range of skills, from methods of instrumenting programs to program optimizations.
To make memory profiling faster, researchers have proposed lossy
techniques to reduce profiling overheads~\cite{vanka:12:ead, chen:04:cc}.
However, such techniques are often of limited utility due to the imprecision
introduced by them.  As one paper puts it, ``\textit{the
difference in accuracy has a considerable impact on the effectiveness of the
speculative optimizations performed}''~\cite{vanka:12:ead}.  Thus, this work
focuses on \textit{precise} memory profiling.  Prior work also proposes
optimizations without sacrificing precision, such as parallelizing the profiler
to reduce the cost~\cite{kim:10:micro}, but these optimizations are often
specific to a particular memory profiler.

Without \textit{practical} memory profilers, memory profiling and its clients like
speculative optimizations are less likely to be adopted.  For
example, Perspective is a state-of-the-art speculative automatic parallelization
system that requires memory profiling~\cite{apostolakis:2020:Perspective}.  To
collect the memory profiles, it uses two off-the-shelf memory profilers, LAMP
and the Privateer profiler~\cite{mason2009lampview,johnson:2012:Privateer}.
Both are state-of-the-art for the memory profiles they produce.  LAMP is a
loop-aware memory dependence profiler that tracks memory dependences and their
loop distances. The Privateer profiler (referred to as ``Privateer'' for
short in this paper) gathers multiple types of memory profiles, including
points-to information, object lifetime, and value predictions.  Both profilers
are based on LLVM, making them easy to integrate into modern compiler
optimizations.  However, their implementation is quite complex, making
them hard to adapt.  They also have significant runtime overhead and fail on
some complex benchmarks.  These problems significantly limit the applicability
of Perspective.

This paper introduces a novel factorization of memory profiling to simplify the
process, enabling developers to focus solely on the core profiling
logic. This approach first separates memory profiling into two main phases:
the frontend and the backend.
\textbf{The frontend} is responsible for generating memory profiling events.
\textbf{The backend} processes these events to produce profiles.
Generalization is then applied to both phases.
The frontend standardizes the instrumentation of common events in memory
profiling while the backend
generalizes and provides commonly used memory profiling
components, such as data structures, algorithms, and optimizations.

Building on this factorization, we present PROMPT, the first memory profiling
framework for \textit{streamlined} development of \textit{fast} memory
profilers. PROMPT systematizes memory profiling events and provides
generalized implementations of both typical profiling components and
optimizations.  Using PROMPT, developers can design and implement memory
profilers more efficiently without delving into compiler internals, parallel
programming, or repeated implementation.
Profilers developed with PROMPT are robust and performant. 
This can shift the perspective on memory profiling adoption. Rather than settling for
off-the-shelf memory profilers with subpar performance  or facing the daunting task of 
building a profiler, developers can now craft tailored memory profilers with low profiling
overhead easily.

This paper offers the following main contributions:
\begin{itemize}
    \item proposes a novel factorization of memory profiling to simplify the development of memory profilers through separating the profiling frontend and backend and generalizing components and optimizations (\S\ref{sec:overview});
    
    \item introduces PROMPT, an open-source, fast, and extensible memory
    profiling framework, and discusses its design and implementation
    (\S\ref{sec:design}, \S\ref{sec:impl});
    
    \item demonstrates the extensibility and performance of PROMPT by porting
    two state-of-the-art memory profilers, LAMP and the Privater
    profiler, and achieving 65\% reduction of the codebase and 5.3$\times$ and 7.1$\times$ faster profiling speed respectively (\S\ref{sec:eval-extensibility}, \S\ref{sec:eval-speed});
    
    \item highlights PROMPT's impact on memory profiling with a
    redesigned memory profiling workflow for a sophisticated compiler optimization 
    client, Perspective, which is succinct at 570
    lines of code and reduces client profiling time by more than 90\%
    (\S\ref{sec:eval-perspective}).
\end{itemize}

%% file: sections/background.tex
\section{Background}
\label{sec:background}

To build an understanding of the functionality and the overhead of memory
profilers from the ground up, we first analyze a typical memory dependence
profiler and discuss the causes of the slowdown.
Then we discuss the workflow of using memory profiling with existing
systems and the difficulties in each option and show how PROMPT changes the
situation of adopting memory profiling.

\subsection{A Typical Memory Profiler}\label{sec:background-typical-memory-profiler}
The memory dependence profiler is the most common type of memory profiler. This
section introduces the design of a typical memory profiler and shows sources of
slowdown. 

\paragraph{Design.}
Consider a vanilla memory dependence profiler that records the set of
manifested memory flow dependences.  A memory flow (i.e., read-after-write)
dependence occurs when a memory load depends on the result of a memory store.
The profiler first instruments the memory instruction and its accessed memory
location for all the memory accesses.  Subsequently, it identifies whether a new
memory access creates a memory dependence.  To do so, the profiler remembers for
each memory address the store instruction that last touches it.  When a load
instruction is executed, a dependence is found from the latest store to the same
memory location as the load instruction.  Dependences, as pairs of load and
store instructions, are then recorded in a data structure.

\paragraph{Slowdown.}
The three steps, namely instrumentation, tracking the latest writes, and
recording dependences in a data structure, all add additional instructions to each
execution of a memory instruction in the original program.  Depending on the
instrumentation method, the added cycles may have various sources, such as
function calls and dynamic translation.  A hash map can be used to track the
latest write to each memory address and a hash set for recording the profiling
results; each comes with additional overhead.  Depending on the implementation,
an additional tens to thousands of CPU cycles can be added to each memory
access, causing an overall slowdown of several to hundreds of times.

% \paragraph{Profiling Task Adjustment.}
% Consider extending the aforementioned memory dependence profiler to record whether a dependence crosses a loop iteration for each loop (i.e., loop-aware).
% Additional events like loop iterations must be instrumented.
% A different logic is also required to keep track of loop contexts as well as the latest write to each memory address.
% A new container, like a map, might be required to capture additional information for loop awareness.
% Thus, even a small addition to the profiling task requires a thorough understanding and modification of an existing profiler.

\addtolength{\tabcolsep}{-4pt}
\begin{figure}[h]
    \begin{minipage}{0.30\textwidth}
        \scriptsize
        \begin{tabular}{ll}
            \toprule
            Category & Systems                                  \\ \midrule
            Compiler & LLVM~\cite{LLVM:CGO04}                   \\
                     & GCC~\cite{gcc}                           \\ \midrule
            \multirowcell{3}{Instrumen-                         \\ tation  \\ System} & Pin~\cite{wallace:07:superpin} \\
                     & DynamoRio~\cite{bruening:2012:dynamorio} \\
                     & Valgrind~\cite{nethercote2007valgrind}   \\ \midrule
            \multirowcell{3}{Memory                             \\ Tracing} & drcachesim~\cite{drcachesim} \\
                     & adept~\cite{zhao:2006:dep}               \\
                     & mTrace~\cite{mtrace}                     \\ \midrule
            \multirowcell{3}{Memory                             \\ Profilers} & SD3~\cite{kim:10:micro}                           \\
                     & LAMP~\cite{mason2009lampview}            \\
                     & Privateer~\cite{johnson:2012:Privateer}  \\ \midrule
            \makecell{Memory                                    \\Profiling\\ Framework}& \textbf{PROMPT (this work)} \\
            \bottomrule
        \end{tabular}
        \vspace{2em}
        \captionof{table}{ Different systems for memory profiling.  }
        \label{tab:related-work}
    \end{minipage}%
    \hfill
    \begin{minipage}{0.62\textwidth}
        \includegraphics[width=\textwidth]{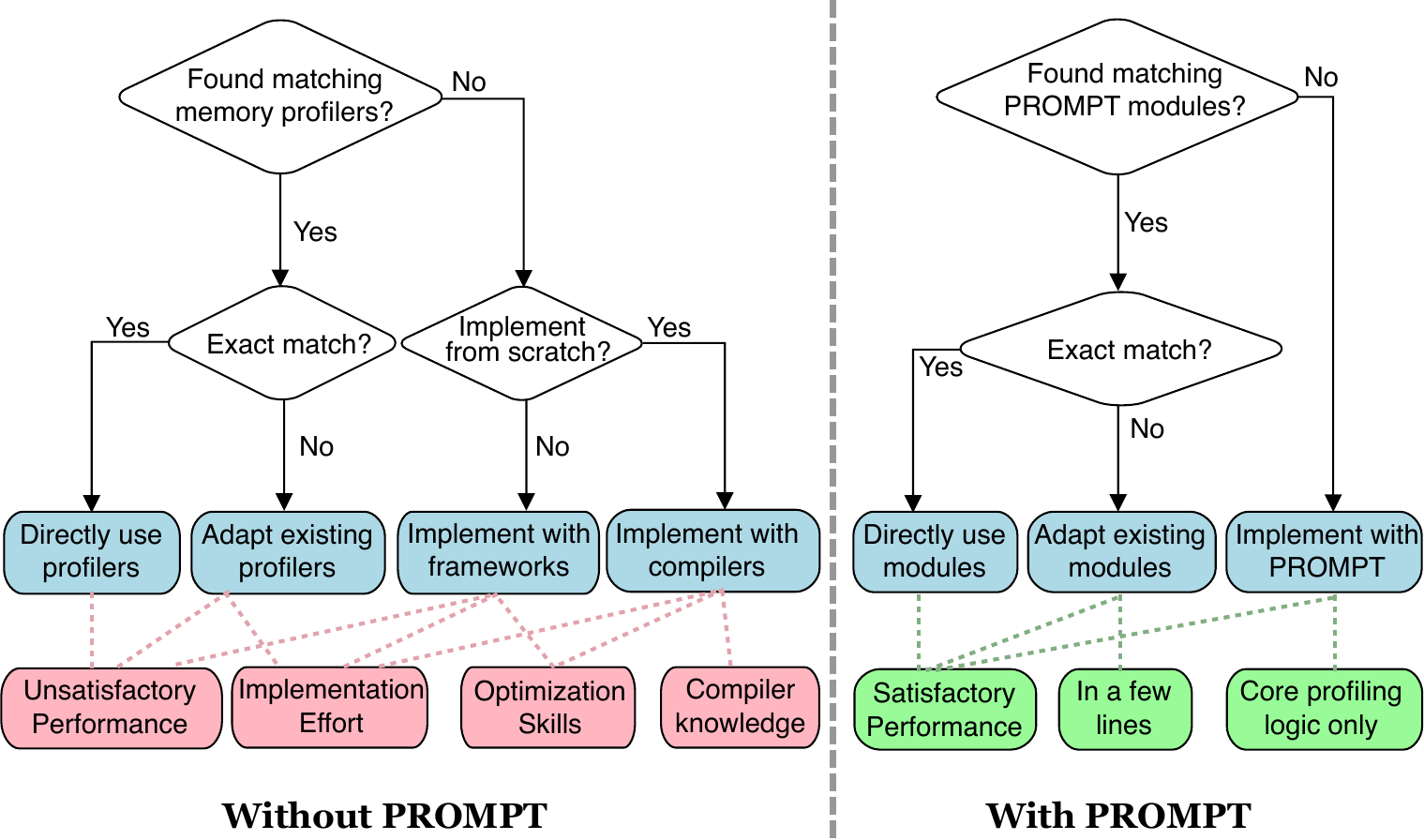}
        \captionof{figure}{The workflow of using memory profiling with and without PROMPT.}
        \label{fig:compare-workflow}
    \end{minipage}
    % \caption{The workflow of using memory profiling with existing systems or PROMPT.}
\end{figure}
\addtolength{\tabcolsep}{4pt}

\subsection{Different Ways to Use Memory Profiling}

Figure~\ref{fig:compare-workflow} illustrates the workflow differences when
using memory profiling with and without PROMPT, based on the systems listed in
Table~\ref{tab:related-work}.

Before PROMPT, when users wish to use memory profiling, the first step is to check
if there are existing profilers matching their requirements.
SD3, LAMP, and Privateer are examples of such existing memory
profilers\cite{kim:10:micro,mason2009lampview,johnson:2012:Privateer}. If the
requirements diverge even slightly, adapting the tool for a new purpose becomes
challenging due to its legacy codebase and monolithic design. Furthermore,
existing memory profilers often have a high overhead, rendering them
impractical. If no suitable profiler exists, users must create a memory
profiler. Instrumentation or memory tracing systems can aid in the development of memory
profilers. While specific compiler knowledge isn't mandatory, significant
implementation effort and optimization skills remain essential. Instrumentation
systems typically operate at the binary level, such as Pin, DynamoRio, and
Valgrind~\cite{luk:2005:pin,bruening:2012:dynamorio,nethercote2007valgrind}.
Dynamic injection of instrumentation code by binary instrumentation systems
leads to an overhead of around 1-10x, irrespective of the profiling logic's
complexity~\cite{luk:2005:pin,nethercote2007valgrind}. Tracing systems, on the
other hand, create and store execution traces, processing them through online or
offline analytical algorithms. For instance, drcachesim,
adept, and mTrace are memory tracing systems built atop
DynamoRio or Pin~\cite{drcachesim,zhao:2006:dep,mtrace}. While these systems
mitigate some implementation efforts, merely gathering the trace incurs a
10--100$\times$ overhead. This does not account for processing the acquired trace
to derive profile information. An alternative approach to building a memory
profiler from the ground up is to leverage a compiler directly, instrumenting at
the intermediate representation (IR) level, as seen with LLVM and
GCC~\cite{LLVM:CGO04,gcc}. Efficient memory profiling can be achieved this way,
as other instrumentation-based systems have shown~\cite{asan,msan}. However,
users should be well-acquainted with the compiler and adept at optimizing
intricate systems.

The memory profiling workflow is simplified with PROMPT. The
initial step involves searching for appropriate modules within the PROMPT
repository. If suitable modules are identified, they can be used directly. If
adaptation is necessary, it typically requires minimal code adjustments, thanks
to PROMPT's modular design. If there is no match and a new implementation is
needed, users can concentrate solely on the profiling logic and delegate the
rest to PROMPT. PROMPT optimizes the workflow and diminishes the
challenges associated with adopting memory profiling.

%% file: sections/overview.tex
\section{Overview: The PROMPT Approach}
\label{sec:overview}

\begin{figure*}[!ht]
    \centering
    \includegraphics[width=\textwidth]{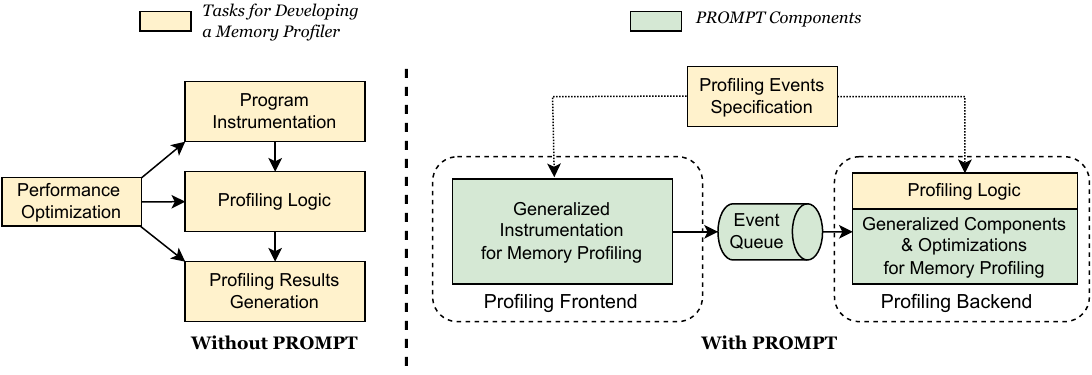}
    \caption{The process of building a memory profiler, with and without PROMPT.}
    \label{fig:prompt-design-overview}
\end{figure*}

As discussed in \S\ref{sec:background-typical-memory-profiler}, a memory
profiling pipeline can be broken down into three steps --- instrumenting
profiling events, the profiler-specific logic that generates profiling results,
and recording and storing profiling results.
Thus, crafting a memory profiler necessitates an intricate understanding
and considerable effort in areas such as instrumentation, the formulation of
profiling logic, and storing the profiling results in certain data
structures—all while ensuring expedient performance, as illustrated on the left
side of Figure~\ref{fig:prompt-design-overview}.
However, it is really the core profiling logic a profiler developer is
interested in.
To allow developers to focus only on the core profiling logic, this paper
presents a novel factorization of memory profiling pipeline, termed as
\textit{the PROMPT approach}.

\paragraph{Separation.}

Inspired by the implementation of some existing
profilers~\cite{ketterlin:2012:profiling,johnson:2012:Privateer,CARMOT},
the PROMPT approach first decouples the profiling into two parts: the event
generation (`frontend') and the profile formulation (`backend').
The profiling frontend instruments the program and tracks profiling events and
corresponding values.  The profiling backend consumes the events, runs the
profiling logic, and generates the profiles.
The frontend and the backend are connected through an event queue. 
This clear separation has three main benefits. 
First, it allows the profiler writer to separate the concerns of instrumentation
from the core profiling logic.  Second, it reduces the interference of profiling
logic with the program under profile.  Moreover, it makes it easier to have
multiple profiling backends to enjoy parallelism without rerunning the program.
While the design of separated frontend and backend has been used in existing
profilers, PROMPT is the first to generalize it as a unified framework that
applies to all existing memory profilers.  The enforcement of decoupled
frontend and backend while easing the connection between them is the key to
PROMPT's extensibility and performance.

\paragraph{Generalization.}
Separation alone doesn't guarantee extensibility and performance. Another
observation is that existing memory profilers have many overlapping components
and optimizations.
By generalizing these components and optimizations, a memory profiler can be
built much more easily and efficiently.
More precisely, the generalization process involves identifying common
components with similar functionalities and developing them with a flexible
interface.  The interface should be easily specialized by the profiler
developers for their specific needs.
For the profiling frontend, the profiling events need to be generalized.
The profiler developers should be able to choose from a set of categorized and
standardized events, and only the events and values the profiler requires will
be instrumented.
Meanwhile, in the profiling backend, shared data structures, algorithms, and
optimizations should be made generic, serving as foundational elements for
developers when implementing their profiling logic.

With the PROMPT approach, the profiler developers only need to specify the
profiling events and implement the core profiling logic, as shown on the right
side of Figure~\ref{fig:prompt-design-overview}.

%% file: sections/design.tex
\section{Design}
\label{sec:design}

\begin{figure*}[t]
    \centering
    \includegraphics[width=\textwidth]{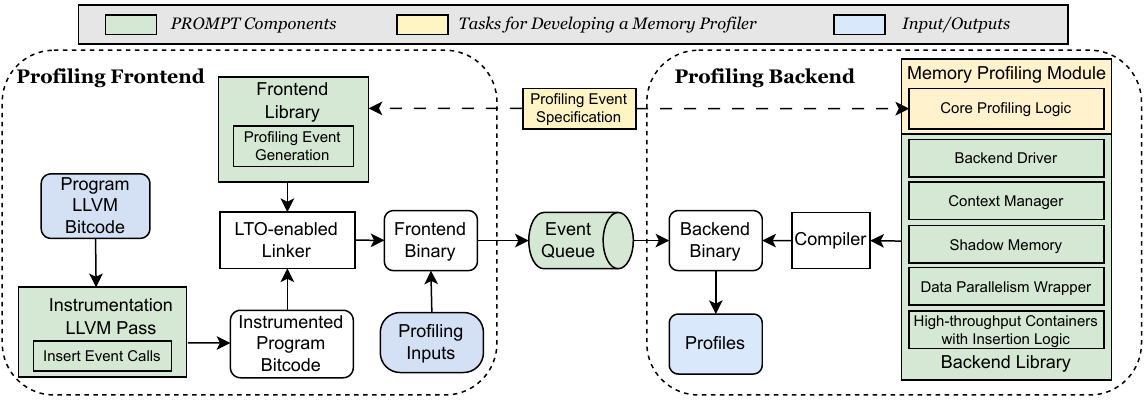}
    \caption{The design of PROMPT.}
    \label{fig:prompt-design}
\end{figure*}

Figure~\ref{fig:prompt-design} shows the design of PROMPT and the workflow
of a memory profiler implemented on top of it.  PROMPT instruments the program
with profiling events in the profiling frontend and generates the frontend
binary.  To implement a memory profiler, one implements a profiling module in
the profiling backend with the help of PROMPT's backend library, and compiles it
to a backend binary.  A profiling process happens when we run the frontend and
backend binary with profiling inputs.  The profiling process and the backend
process communicate through the event queue.  The backend process will generate
the profile.

\subsection{Generalizing Memory Profiling Components}
\label{sec:design-components}
The profiler writer still needs to implement many functionalities to build a
memory profiler, many of which are common across different memory profilers.
For example, many profilers need to keep a map from the memory address to the
metadata.
PROMPT recognizes this and provides a set of common components to ease the
implementation of the logic of a memory profiler.

\paragraph{Profiling Frontend}
PROMPT introduces a generic frontend designed to instrument the program, thereby
facilitating the generation of memory profiling events. A categorization and
standardization of profiling events, prevalent in existing memory profilers, is
performed. Moreover, each event encompasses a set of arguments. Section
\S\ref{sec:impl-events} discusses the event types and their respective arguments
in detail. Additionally, PROMPT instruments the source code with callback
functions, which sequentially push the profiling events to the queue.

\paragraph{Profiling Backend}
PROMPT provides an array of generic backend components to streamline the
development of memory profilers. Shadow memory profiling, previously employed in
various dynamic program analysis tools~\cite{zhao:10:ismm,nethercote:07:vee},
operates by storing metadata in a distinct shadow memory location. PROMPT
includes a versatile shadow memory that can be tailored to accommodate specific
metadata requirements. Often, memory profilers necessitate tracking context,
such as the call stack and loop nest, with the context information of a
particular event potentially being encoded in the shadow memory for future
retrieval. PROMPT provides a generic context manager adept at encoding and
decoding such contexts. Furthermore, memory profilers frequently utilize
containers, such as sets and maps, coupled with a certain insertion logic to
document profiling results — for instance, generating a new entry or
incrementing a count in a map from dependence. PROMPT offers various containers
equipped with predefined insertion logic to facilitate this process.

\subsection{Generalizing Memory Profiling Optimizations}
Optimizations are imperative for memory profilers to ensure viable performance
and practical utility. While numerous optimizations are prevalent across
existing memory profilers, their instantiation in disparate forms renders
generalization a nontrivial task. PROMPT facilitates a generalized approach to
two main optimizations, specialization and parallelism, thereby enabling most
memory profilers to use them with minimal developmental effort.

\paragraph{Removing unnecessary instrumentation.}
\label{sec:design-tailoring}

Memory profilers may only care about a subset of events.
We use the specialization technique to remove unnecessary events and reduce overhead~\cite{reps:1996:programSpecialization}.
A way to do specialization can be at the instrumentation time.
We can configure the LLVM pass to only instrument the necessary calls and arguments.
However, this requires a complicated way to communicate with the LLVM pass.
Instead, PROMPT does the specialization at link-time.
As shown in Figure~\ref{fig:prompt-design}, PROMPT gets the profile event specification from the module implemented by each profiler.
PROMPT then automatically specializes the frontend library that generates profiling events to the queue based on the specification.
For any irrelevant event, an empty function body will be generated.
For any information not required for an event, the argument will not be pushed to the queue.
% All the callback functions are marked as \texttt{inline}.
At link time, we enable link-time optimization. The compiler will automatically
optimize any dead instructions, empty functions, unused arguments, and all
instrumented code to produce them (see profiling frontend in
Figure~\ref{fig:prompt-design-overview}).  In this way, PROMPT removes the cost
introduced by generic events without configuring the LLVM pass.  We have
verified the validity of this approach by examining the generated binaries to
confirm that the generic event handling was removed.
This link-time specialization makes the instrumentation LLVM pass easy to
implement and easy to maintain.

\paragraph{Data Parallelism}
\label{sec:design-generalize-parallelism}

Another common optimization among memory profilers is the parallelism.
PROMPT makes it easier to leverage parallelism.
One form is the address-based parallelism that state-of-the-art memory
profilers implement for their specific task~\cite{kim:10:micro}.  Multiple
profiling backends can run in parallel to process profiling events to decoupled
chunks of address space, as shown in the profiling thread of
Figure~\ref{fig:prompt-design-overview}.
PROMPT generalizes this to other types of data parallelism, such as parallelism
of tasks on different originating instructions instead of different addresses.
It also provides a wrapper to adopt data parallelism easily. 
A memory profiler built with PROMPT
only needs to mark that an operation is decoupled based on the address of other
values and provide a method for merging results.  PROMPT will manage the
parallelism at runtime.

\subsection{Trading Latency for Throughput}

PROMPT uses a pivotal insight to enhance performance markedly: trading
latency for throughput. Here, throughput is defined as the number of events
processed within a given time unit, while latency represents the time interval
between a memory event's generation at the frontend and its processing at the
backend. Given that memory profilers only supply aggregated summaries—or
\textit{profiles}—upon completion and do not necessitate real-time feedback,
latency does \textbf{not} emerge as a critical aspect for memory profilers.
However, due to the typically immense data volumes generated in memory
profiling, a system with high throughput becomes imperative to expediently
process the memory events. Any bottleneck in the queue, profiling logic, or
result-storing containers will hamper the entire process. Consequently, numerous
components within PROMPT are intentionally crafted to prioritize throughput over
latency.

This optimization is primarily realized by incorporating buffers into
bottleneck-inducing components, thereby redistributing the load to other
components which can harness parallelism or alternative optimizations to boost
throughput.
One example is the queue situated between the frontend and backend. We
identified the main throughput bottleneck as the overhead of
writing events to the frontend queue. PROMPT counteracts this by employing a
blend of ping-pong buffer design and streaming writes, ensuring the frontend can
inscribe to the queue with minimal latency (refer to \S\ref{sec:impl-queue} for
further details). Another illustration involves the containers responsible for
storing profiling results. It is customary for these containers to experience a
deluge of stores within a brief window, interspersed with periods devoid of
reads. Thus, PROMPT utilizes a buffer to aggregate the stores, performing the
reduction (typically in parallel) only when the buffer reaches capacity or when
a read is initiated (see \S\ref{sec:impl-containers} for additional details).

%% file: sections/implementation.tex
\section{Implementation}
\label{sec:impl}

PROMPT's frontend, backend, and profiling modules are developed in C++, while
its instrumentation is built upon the LLVM compiler infrastructure.  
PROMPT's instrumentation pass is currently built on LLVM 9.0.1 in order to align
with the latest versions of LAMP, Privateer, and
Perspective~\cite{cpf:github}. 
The frontend has around 3400 lines of code, with 2600 for the instrumentation
pass and 800 for the frontend library.  The backend has around 3000 lines of
code, with 900 for the backend driver, 400 for the context manager, 200 for the
shadow memory, 100 for the data parallelism wrapper, 1000 for the
high-throughput data structures, and some other utilities. The queue has around 500
lines of code.
Note that PROMPT is still under active development, so the code size may change
in the future.
PROMPT is open-source~\cite{prompt_NON_ANON_XXXX}.

\subsection{Profiling Events}
\label{sec:impl-events}

\begin{table}[!ht]
    \small
    \centering
    \begin{tabular}{ccc}
        \toprule
        Event  Category                & Events                & Information                           \\ \midrule
        \multirow{3}{*}{Memory Access} & Load                  & Instruction ID, address,  value, size \\
                                       & Store                 & Instruction ID, address, value, size  \\
                                       & Pointer Creation      & Instruction ID,  address, type        \\ \midrule
        \multirow{7}{*}{Allocation}    & Heap Allocation       & Instruction ID, address, size         \\
                                       & Heap Deallocation     & Instruction ID, address               \\
                                       & Stack Allocation      & Instruction ID, address, size         \\
                                       & Stack Deallocation    & Instruction ID, address               \\
                                       & Global Initialization & Object ID, address,  size             \\ \midrule
        \multirow{7}{*}{Context}       & Function Entry        & Function ID                           \\
                                       & Function Exit         & Function ID                           \\
                                       & Loop Invocation       & Loop ID                               \\
                                       & Loop Iteration        & Loop ID                               \\
                                       & Loop Exit             & Loop ID                               \\
                                       & Program Starts        & Process ID                            \\
                                       & Program Terminates    & Process ID                            \\
        \bottomrule
    \end{tabular}%
    \caption{The profiling events provided by PROMPT.}
    \label{tab:profiling-events}
\end{table}

PROMPT provides three categories of profiling events --- memory access, allocation, and context events, as listed in Table~\ref{tab:profiling-events}.
Most events are instrumented at the LLVM IR level by adding callback functions right after the corresponding event with all the information through function arguments.
For example, a load event will be followed by an \texttt{onLoad(instrId, address, value, size)}.
% Heap allocation events are an exception.
%
Heap allocation and deallocation events are tracked using library
interposition, so allocations in external functions can be tracked to provide a
complete view of the memory space~\cite{hooks}.

% With the existence of external functions that do allocation inside, only tracking calls to allocation functions in LLVM IR is sometimes not enough.
% With hooks, all memory allocation events can be tracked.

\subsubsection{Adding Profiling Events}
PROMPT provides a comprehensive set of profiling events, adequately addressing
the requirements of many existing memory profilers, yet the necessity for
additional events in future developments is acknowledged. Although the addition
of new events presents its own challenges, the decoupled design of PROMPT
facilitates a clearer and more straightforward implementation process compared
to current memory profilers. The procedure involves initially specifying the
event and its potential values, followed by crafting the instrumentation in the
LLVM pass, and finally integrating the corresponding callback function into the
frontend library. It is noteworthy that designing the instrumentation is the
most complex part of this process, requiring a solid understanding of the LLVM
IR.

\subsection{Event Queue}
\label{sec:impl-queue}

\begin{figure}[!ht]
\centering
    \includegraphics[width=0.8\columnwidth]{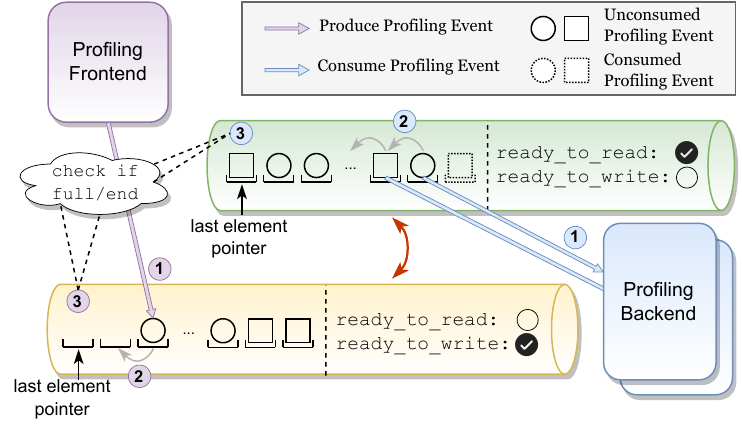}
    \caption{High-throughput SPMC Queue}
    \label{fig:spmc-queue}
\end{figure}

The queue helps PROMPT break up the frontend and backend.
PROMPT needs an SPMC (single-producer-multiple-consumer) queue, where the
producer is the profiling frontend, and the consumers are multiple workers in
the profiling backend.  We observe that memory profilers do not have latency
requirements for the queue.  The additional cycles introduced by the
instructions instrumented rather than the memory throughput bounds the queue
performance~\cite{jablin2010epic}.  We implement a high-throughput SPMC queue
specialized for the memory profiling task.
The queue uses a ping-pong buffer design~\cite{swaminathan:2012:pingpong} as
shown in Figure~\ref{fig:spmc-queue}.
The advantage of a ping-pong buffer design lies in the fact that producers and
consumers do not need to communicate until one buffer reaches its capacity.
Thus, the producer can keep writing to one buffer, without communication, until
that one is full. Then, it checks whether the other buffer is ready.  The vice
versa is true for the consumer.  This greatly reduces the communication overhead
between the producer and the consumer.

To reduce the wait time of the writes and reduce interfering with the program
under profiling, the queue uses streaming
writes\cite{krishnaiyer:2013:streaming}.  Streaming write is a feature of the
X86 architecture. It bypasses the cache hierarchy and improves the frontend code
performance by avoiding ``contaminating'' the cache.  The writes are very
efficient by using a relatively large buffer (more than 1MB). 

The SPMC queue is bounded, thus the producer and consumers must communicate at
the end of one buffer by checking whether the other buffer is ready.  We can
reduce the frequency of checking by making the buffer bigger, leveraging the
latency-insensitive insight.  A bigger buffer also makes parallelism at the
backend more efficient by amortizing the cost of parallel workers.  With
streaming writes, the buffer already bypasses the cache hierarchy, so a bigger
buffer size has minimal performance drawbacks.  The large buffer size also
smooths out spikes in the producer.

\subsection{Backend Components}

\paragraph{Backend Driver.}
The backend driver consumes the events from the queue and calls the corresponding call of the profiling modules.
It manages profiling threads if data parallelism is used (\S\ref{sec:design-generalize-parallelism}).

\paragraph{A Generic Shadow Memory.}
PROMPT provides a generic shadow memory that can be configured to fit metadata
of different sizes.  It takes care of allocation and deallocation automatically.
PROMPT applies a direct mapping scheme that applies a shift and mask to all
memory addresses to translate from program to shadow addresses.
It is an efficient implementation of the map from the memory address to the
metadata. 

\paragraph{A Generic Context Manager.}
The context manager in PROMPT provides a generic way to manage the context.  It
interacts with a profiler to transform, encode, and decode a context.  It keeps
track of the current context through transform APIs (e.g.,
\texttt{pushContext(type, ID), popContext(type, ID)}).  It provides multiple
ways to encode and decode a context.  One way is through a map of manifested
context to a counter.  Caching optimizations are used to reduce the lookup cost
of decoding the context.  If the context is simple enough, the context manager
will use the concatenation of the context as the encoding.  Note that due to
synchronization, sharing one context manager can be problematic, so PROMPT
maintains a separate context manager for each backend thread.

\begin{figure}[h]
    \centering
    \includegraphics[width=0.70\columnwidth]{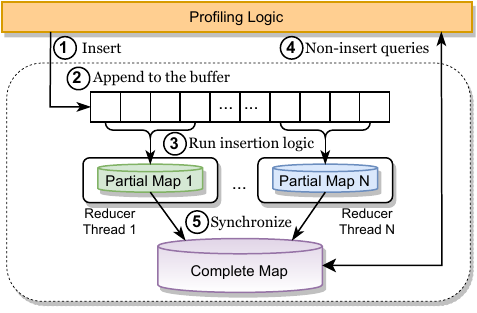}
    \caption{A high-throughput hash map in action.}
    \label{fig:reducible-container}
\end{figure}

\paragraph{Data Structures with Insertion Logic.}
\label{sec:impl-containers}
To help
memory profilers simplify the logic, PROMPT provides data structures with
built-in insertion logic, including checking for constant, counting,
summing, or finding the minimum and maximum.
\texttt{htmap\_constant} is a map from a key to
the value if it is constant.
\texttt{htmap\_count} is a map from a key to its
count.
\texttt{htmap\_sum/min/max} is a map from a key to the sum, minimum, or
maximum of all values corresponding to a key.
\texttt{htmap\_set} is a map from a value to a set with an optional size limit.
One thing in common with all these data structures 
is that the insert operation is reducible.  For example, for \texttt{Map\_Sum},
which provides a map from keys to the sum of values, inserting to this map
translates to summing up values to each key, which is a reducible operation. A
reducible operation can be executed in any order in parallel.
With this observation, we provide parallelism as a part of these maps.  As
shown in Figure \ref{fig:reducible-container}, all the insertion to the
map is buffered to a vector with a fixed reserved size, and once the
buffer is full, many workers will do the reduction in parallel.  Each takes a
chunk of the buffer and reduces it to its local map.  Only when any API
other than insertion is called will the workers merge the local map into
the global one.  This design works well with a memory profiler, where insertion
is almost always the only operation to the profiling data structure during profile
time.  To improve efficiency, PROMPT adopts a thread pool, where the reduction
thread will stay in the background waiting for the tasks.
In addition to the maps, PROMPT also provides drop-in
replacements for \texttt{set} and \texttt{unordered\_set(hash set)} that provide
the same optimization. These replacements do not offer complete C++ STL support;
however, they handle common APIs that adequately meet the requirements of a
memory profiling module.

\begin{listing}[t]
    \input{sections/loaded-value-example-code.tex}
    \caption{The implementation of a value pattern profiler that checks for
        constant loaded values.  }
    \label{code:loaded-value}
\end{listing}

\subsection{Implementing a Memory Profiler with PROMPT}

\label{sec:impl-profilers}
% \todo[inline]{Discussion of how to build a memory profiler with the components}

To implement a new memory profiler with PROMPT, one only needs to declare the
subset of relevant events listed in Table~\ref{tab:profiling-events} and
implement core profiling logic in the callback functions of events.  The
components defined in \S\ref{sec:design-components} and
\S\ref{sec:design-generalize-parallelism} are available to ease the
implementation and provide good performance out of the box.
Listing~\ref{code:loaded-value} shows how a value pattern profiler works. It
tracks all loads and records those that have constant loaded values.

Here are several example memory profilers implemented with PROMPT and the core
logic.
A \textbf{Memory Dependence Profiler} tracks the source and destination of a
memory dependence, optionally the related loops, contexts, and counts.  A
profiler can use shadow memory to track the last load/store instruction and
additional information to each memory address, the record dependence if
discovered.  Figure~\ref{fig:prompt-design-overview} shows a memory-dependence
profiler.
A \textbf{Value Pattern Profiler} tracks whether the value of a memory access
follow some patterns, such as always a constant.  A profiler can use PROMPT's
components to automatically check for the constant pattern. The module in
Listing~\ref{code:loaded-value} shows such a profiler.
A \textbf{Points-to Profiler} maps each pointer to the set of memory objects
that it points to.  A profiler can first uniquely identify memory objects at
allocation time using the instruction ID and the context tracked by the context
manager, track the object information in the shadow memory, and record the
object associated with the specific address at pointer creation time.
An \textbf{Object-Lifetime Profiler} tracks the lifetime of each object and
checks if it is dynamically local to a scope such as a loop.  A profiler can
track the uniquely identified memory objects similar to  the points-to profiler,
check the shared context of allocation and deallocation, and record the object
and the shared context.

%% file: sections/loaded-value-example-code.tex
\begin{minted}[
framesep=2mm,
baselinestretch=1,
fontsize=\small]
{yaml}
# Profiling events specified in YAML
module: ValuePatternConstantLoadModule
events: # and the corresponding values
  load: [instruction_id, value]
  finished: []
\end{minted}
\rule{\linewidth}{0.4pt}
\begin{minted}[
framesep=2mm,
baselinestretch=1,
fontsize=\small]
{C++}
// Core profiling logic
class ValuePatternConstantLoadModule : public DataParallelismModule,
                                       public ProfilingModule {
private:
  // High throughput map provided by PROMPT that checks if the value is constant
  HTMap_Constant<InstrId, LoadedValue> constmap_value;

public:
  // The `num_threads` and thread id (`tid`) are used to control the data parallelism.
  // They are automatically set by the driver on initialization.
  LoadedValueModule(uint32_t num_threads, uint32_t tid) :
                DataParallelismModule(num_threads, tid) {}
  // On every Load event, the instruction ID and value are passed in.
  void load(uint32_t instrId, uint64_t value) override {
    // A wrapper by DataParallelismModule:
    // This will only execute if the worker is in charge of the instruction ID.
    execute_if_mine(instrId, [&]() {
      // insert the ID and value to the map
      constmap_value.insert({instrId, value});
    });
  }

  void finish(string filename) override {
    // Dump the constmap_value in a format required by the client
  }

  // When using data parallelism, need to implement how modules are merged.
  void merge(LoadedValueModule &other) override {
    // merge the map from instruction ID to value
    constmap_value.merge(other.constmap_value);
  }
};
\end{minted}

%% file: sections/evaluation.tex
\section{Evaluation}
\label{sec:eval}

PROMPT is designed for extensive extensibility, seamlessly supporting a wide
array of applications. As \S\ref{sec:eval-extensibility} illustrates, porting
LAMP and Perspective, two state-of-the-art LLVM-based profilers, to PROMPT
reduces more than half of the code size and makes the code easier to understand.
Many variants of memory dependence profilers can also be adapted from a basic
profiling module with a few lines of code.

In terms of speed, evaluations in \S\ref{sec:eval-speed} compare it against LAMP
and the Privateer profiler on SPEC CPU 2017 benchmarks, showing that PROMPT is
5\Times{} faster than LAMP and 6\Times{} faster than Privateer profiler on
average.  Moreover, across a myriad of memory dependence profilers with diverse
goals and technologies, PROMPT's speed is consistently equivalent or superior
(\S\ref{sec:eval-mem-dep}).

\S\ref{sec:eval-perspective} underscores the impact of PROMPT by redesigning the
memory profiling workflow for Perspective~\cite{apostolakis:2020:Perspective}.
In 570 lines of code, the new workflow satisfies Perspective's memory profiling
needs while reducing profiling overhead by 95\%.  The new workflow is also more
applicable to more complex applications.
The design elements of PROMPT are evaluated separately to understand how they
drive PROMPT's performance in \S\ref{sec:eval-perf-analysis}, where PROMPT's
memory and binary size overheads are also discussed.

\subsection{Experiment Setups}
\label{sec:eval-setup}

All performance experiments are run on a machine with two Intel Xeon E5\-2697 v3
processors with 252~GB of memory.  The operating system is 64-bit Ubuntu 20.04
LTS\@. 

PROMPT is evaluated against SPEC CPU 2017 suite when comparing against LAMP and
Privateer.  Each benchmark is first
compiled and linked into one LLVM bitcode file, which
is the same preprocessing workflow as LAMP and Privateer, and required by Perspective.  Due to the
limitation of this pipeline, FORTRAN benchmarks (lack of \texttt{flang} for the
LLVM version) and \texttt{502.gcc} (\texttt{muldefs} not supported with
\texttt{llvm-link}) from SPEC 2017 are excluded.  The evaluation contains 15
C/C++ benchmarks from the SPEC CPU 2017 suite with 3.6 million lines of code
combined~\cite{bucek:2018:spec2017}, In the case study
(\S\ref{sec:eval-perspective}), benchmarks from the Perspective paper are also used
to do the performance comparison~\cite{apostolakis:2020:Perspective}.  All
evaluation use the training inputs since reference inputs would be more
appropriate for evaluating the clients' performance with the profiling
information.

\subsection{PROMPT's Extensibility}
\label{sec:eval-extensibility}
\S\ref{sec:impl-profilers} shows concretely how easily the memory
profilers are implemented.
Memory profilers can be implemented with PROMPT by expressing only the core
logic. Adaptation of existing memory profilers is also much easier with PROMPT.

\begin{table}[h]
\small
\centering
\begin{tabular}{ccc}
    \toprule
    \multirow{2}{*}{Components} & \multicolumn{2}{c}{LOC}                                   \\ \cmidrule{2-3}
                                & Original LAMP           & Ported with PROMPT              \\ \midrule
    Instrumentation             & 713                     & N/A (provided by PROMPT)        \\
    Event Generation            & 803                     & N/A $\ast$ (provided by PROMPT) \\
    Event Specification         & N/A                     & 13                              \\
    Core Profiling Logic        & 668                     & 898$\ast$                       \\
    Memory Map (Shadow Memory)  & 691                     & N/A (provided by PROMPT)        \\ \midrule
    \textbf{Total LOC}          & \textbf{2875}           & \textbf{911}                    \\
    \bottomrule
\end{tabular}
\caption{The comparison of lines of code (LOC) of LAMP and the ported version
with PROMPT\@.  $\ast$Original LAMP does not use the frontend-backend design, so
the event generation directly calls other functions in the core profiler logic.
Thus, the core profiling logic in the ported LAMP subsumes part of the
event generation.
    }
    \label{tab:lamp-loc}
\end{table}

\begin{table}[h]
\small
\centering
\begin{tabular}{ccc}
    \toprule
    \multirow{2}{*}{Components} & \multicolumn{2}{c}{LOC}                            \\ \cmidrule{2-3}
                                & Privateer Profiler      & Ported with PROMPT       \\ \midrule
    Instrumentation             & 3161                    & N/A (provided by PROMPT) \\
    Event Generation            & 464                     & N/A (provided by PROMPT) \\
    Event Specification         & N/A                     & 19                       \\
    Queue                       & 227                     & N/A (provied by PROMPT)  \\
    Core Profiling Logic        & 1401                    & 1486 $\ast$              \\ \midrule
    \textbf{Total LOC}          & \textbf{5253}           & \textbf{1505}             \\
    \bottomrule
\end{tabular}
\caption{The comparison of lines of code (LOC) of the Privateer Profiler and the
ported version with PROMPT.  $\ast$The core profiling logic in the ported
version includes some additional interfacing with the backend driver.  }
\label{tab:privateer-loc}
\end{table}

\paragraph{PROMPT allows developers focus on core profiling logic only}

Two existing memory profilers, LAMP and the Privateer profiler are ported to
PROMPT.  Table~\ref{tab:lamp-loc} and Table~\ref{tab:privateer-loc} show the LOC
of the original and the ported version of them.  \texttt{cloc} tool is used to
count
the lines of code (LOC) Blank lines and comments are excluded~\cite{adanial:2021:cloc}.  
For both profilers, porting to PROMPT reduces around 70\% of the LOC by focusing
on the core profiling logic.
The instrumentation with LLVM alone requires thousands of lines.  Other shared
components like the event generation, shadow memory, and the queue are also
provided by PROMPT to reduce the implementation effort.

\paragraph{PROMPT is easy to adapt}

\begin{table}[ht]
    \small
    \centering
    \begin{tabular}{cc}
        \toprule
        Extensions (incremental)                                              & LOC Delta \\ 
        \midrule
        + Dependence count~\cite{mason2009lampview, ketterlin:2012:profiling} & 1         \\
        + All dependence types~\cite{morewFelix:2020:skipping}                & 10        \\
        + Dependence distance~\cite{yu:2012:fld,kim:10:micro}                 & 7         \\
        + Context-aware~\cite{CAMP,chen:04:cc, sato:2012:whole}               & 16        \\
        \bottomrule
    \end{tabular}
    \caption{The comparison of different variants of the memory dependence
        profilers.  We incrementally extend the memory dependence profiler built
        with PROMPT and present the LOC delta between every two variants.
    }\label{tab:memory-dep-loc-comparison}
\end{table}

The memory dependence profiler is the most well-studied memory profiler.
A memory dependence profiler can track the sources and sinks of
memory dependences, the frequencies, loop-carried or loop-independent,
distances, and contexts, for all types of memory dependences (flow, anti, and
output)~\cite{kim:10:micro,chen:04:cc}.
In Table~\ref{tab:memory-dep-loc-comparison}, we start from a basic
memory flow-dependence profiler and incrementally adapt it to other variants of
memory-dependence profilers by changing a few lines.

\subsection{PROMPT's Speed}
\label{sec:eval-speed}

\subsubsection{Comparing Against LAMP and Privateer Profiler}
\label{sec:eval-speed-llvm}

\begin{figure}[h]
    \includegraphics[width=0.95\columnwidth]{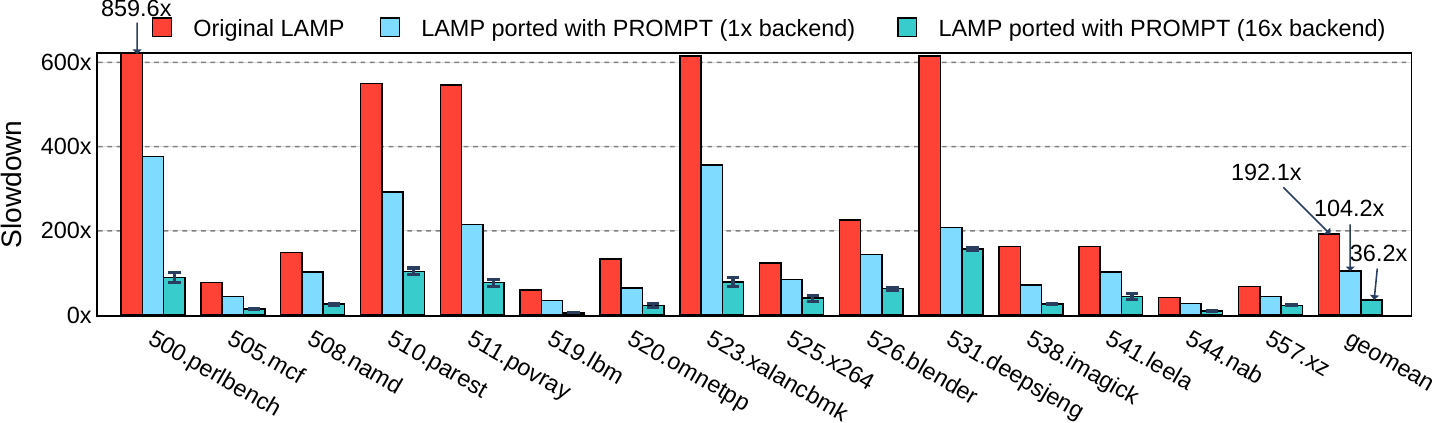}
    \vspace{0.5cm}
    \includegraphics[width=0.94\columnwidth]{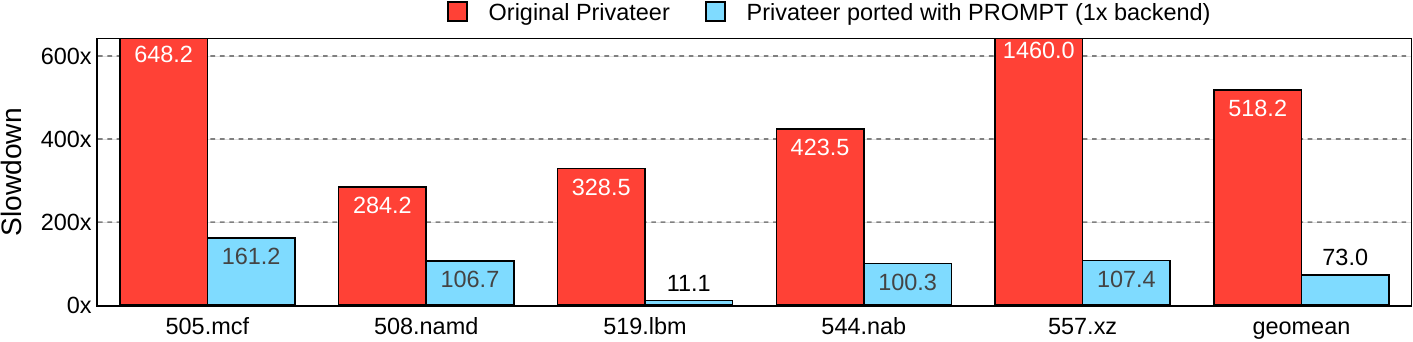}
    \caption{The performance comparison of the original profilers and the versions ported with PROMPT.}
    \label{fig:perf-lamp-privateer-spec2017}
\end{figure}

To ensure a direct and meaningful comparison, LAMP and Privateer, both of which
target LLVM IR—precisely where PROMPT operates, are evaluated.
PROMPT is set to generate equivalent profiling information as the original
profilers and is evaluated on the same set of benchmarks.
We ran each benchmark for original LAMP and Privateer profiler once due to the
long profiling time (more than 10 hours for a few benchmarks).
For all ported version, the data represents the average (mean) of five
runs. The error bars indicate the 99\% confidence interval.
Given that the error bars for all other versions are visually negligible, we
only display the error bars for the ported LAMP with 16 backend threads.
As Figure~\ref{fig:perf-lamp-privateer-spec2017} shown, for LAMP, the ported
version running with 16 threads on the backend runs 5.3\Times{} on average.
The performance improvement first comes from the pipeline parallelism from the
decoupled design. As shown with the ported LAMP with one backend thread,
the performance almost doubles.  The second source of performance improvement
is the parallelism wrapper added in a few lines of code
(\S\ref{sec:design-generalize-parallelism}).  In this experiment, we used up to
16 backend threads to consume the profiling events which brings an additional
three times speedup.

Due to the design limitations, the Privateer profiler fails to run or times out
after 24 hours on ten out of the 15 SPEC 2017 benchmarks. Thus, we compare the
performance on rest five benchmarks in
Figure~\ref{fig:perf-lamp-privateer-spec2017}.  The ported version is
7.1\Times{} faster on average. 
This improvement was above our expectation.  Due to the complex design of the
Privateer profiler, we did not apply data-level parallelism to it.  Moreover,
because the original profiler also has a frontend-backend design, the algorithms
for the original and the ported Privateer profiler are essentially the same.
Upon close inspection, we found that the performance improvement comes from the
optimizations in PROMPT's event queue.  The original Privateer profiler's
bottleneck is in the frontend, during generating events to the queue. PROMPT
significantly reduces the overhead of generating events to the queue using
the high-throughput queue (\S\ref{sec:impl-queue}).

Note that for both ported profilers, we did not alter the core logic or the 
profiling needs to achieve the performance improvement.  The performance
improvement comes from the generalized optimizations in PROMPT.  In
\S\ref{sec:eval-perspective}, we show how to further improve the performance by
redesigning the memory profilers with PROMPT, where we tailor the memory
profiling workflow to the client's needs.

\subsubsection{Comparing Against Other Memory-Dependence Profilers}

\label{sec:eval-mem-dep}

\begin{table}[ht]
    \small
    \centering
    \begin{tabular}{ccc}
        \toprule
        \makecell{Extensions\\ (incremental)}           & \makecell{PROMPT's Geomean Slowdown \\Running on SPEC 2017 Benchmarks} & \makecell{Prior Reported \\Slowdowns} \\ \midrule
        Dependence count     & 7.5\Times{}       & 88--118\Times{} \\ \midrule
        All dependence types & 10.2\Times{}      & 28--36\Times{}  \\ \midrule
        Dependence distance  & 10.8\Times{}      & 5--29\Times{}   \\ \midrule
        Context-aware        & 13.1\Times{}      & 39--132\Times{} \\
        \bottomrule
    \end{tabular}%
    \caption{Comparison of PROMPT's slowdowns against different
    memory-dependence profilers.  PROMPT's slowdowns are the geomean of SPEC CPU
    2017 benchmarks.  Prior work slowdowns are taken from the original papers.
    }
    \label{tab:memory-dep-overhead-comparison}
\end{table}

Table~\ref{tab:memory-dep-overhead-comparison} shows the slowdown of PROMPT
running on SPEC 2017 benchmarks and the slowdown of prior work reported in the
paper.
All other work focusing on memory dependence profiling, listed in 
Table~\ref{tab:memory-dep-loc-comparison}, is implemented with different
technologies and evaluated on different benchmarks on different machines.
Comparing the performance of PROMPT against all these memory profilers is
challenging. Thus, the goal here is to show that the performance of PROMPT is
consistent with existing memory-dependence profilers.
% We want to clarify a technical difference in our implementation of
% memory-dependence profilers.  Some existing memory-dependence profilers gather
% profiling for multiple loops in a single
% run~\cite{kim:10:micro,mason2009lampview}.  In contrast, our implementation of
% memory-dependence profilers focuses on the run for the hottest loop.  A similar
% comparison was used in prior work~\cite{yu:2012:fld}.
% It is valid for two reasons.
% First, the profiling for the hottest loop is the critical path; if needed, the client can run multiple loops simultaneously with task parallelism.
% Second, the hottest loop usually accounts for more than 95\% of the program execution time and all dependences during the loop execution are tracked; thus, the performance difference between tracking the hottest loop and all loops in a program is not several times but several percent. 
% %
% While it is not due to a design limitation of PROMPT, we choose to focus on one loop at a time for all profilers. 
% It makes the memory profilers much simpler and easier to integrate with use cases like automatic parallelization.
% % We intend to keep this design unless motivated by another use case.

\subsection{Redesigned Memory Profiling for Perspective}
\label{sec:eval-perspective}

In the current implementation, Perspective uses LAMP and Privateer Profiler.
In \S\ref{sec:eval-speed-llvm}, PROMPT is evaluated against LAMP and the
Privateer profiler while reproducing their profiling output.
However, not all profiling functionalities and configurations are necessary to
fulfill the profiling needs of Perspective.
With PROMPT, we redesigned the memory profilers to exactly match Perspective's
needs with first principles and show the benefits of PROMPT in this case study.

\begin{table}[ht]
    \small
    \centering

    \begin{tabular}{ccc}
        \toprule
        Profiling Needs                    & PROMPT Profiling Module          & LOC \\\midrule
        Memory Flow Dependence Speculation & Memory  Dependence               & 136 \\ 
        Value Speculation                  & Value  Pattern                   & 69  \\
        Short-lived Object  Speculation    & Object  Lifetime                 & 117 \\
        Points-to Speculation              & Points-to                        & 248 \\ \midrule
        \textbf{Total LOC}                 & \multicolumn{2}{c}{\textbf{570}}       \\
        \bottomrule
    \end{tabular}
    \caption{The lines of code (LOC) of the PROMPT profilers for each profiling
    need of Perspective.}
    \label{tab:PROMPT-and-perspective-loc-comparison}
\end{table}

\paragraph{Streamlined Development}
We first identify the four memory profiling needs of Perspective as shown in
Table~\ref{tab:PROMPT-and-perspective-loc-comparison}.  Because Perspective
works on a loop basis, the memory profilers are only needed for the hottest loop
identified by the compiler.  We implement four memory profiling modules,
memory-dependence profiler, value-pattern, object-lifetime, and points-to
profiling to cover Perspective's needs.  The four memory profilers with PROMPT
only require 570 lines of code, a dramatic reduction of the required
implementation effort.

\begin{figure*}[!ht]
    \centering
    \includegraphics[width=\textwidth]{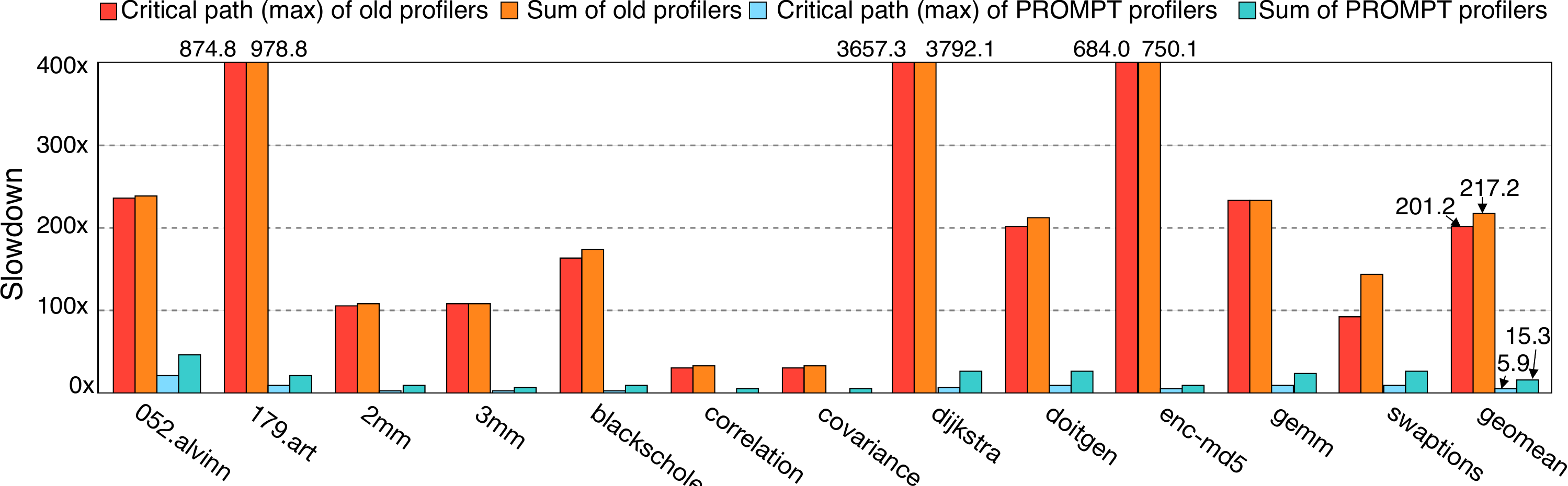}
    \caption{The profiling slowdowns of the existing memory profilers and the
        PROMPT ones on benchmarks from the Perspective paper.}
    \label{fig:e2e-comparison}
\end{figure*}

\paragraph{Faster Profiling.}

We compare the profiling time overhead with PROMPT and the existing profilers
used in Perspective.  The critical path of the profiling workflow is the
longest-running profiler because independent profilers can be executed in
parallel with the same input.  We show the
critical path of both workflows and also the sum of all profilers in each
workflow in Figure~\ref{fig:e2e-comparison}. 
PROMPT reduces the critical path slowdown from 217.2\Times{} to only 5.9\Times{}
and the sum of profiling time from 201.2\Times{} to 15.3\Times{}.
All results are the average (mean) of five runs. 
In our experiments, the maximum coefficient of variation (the ratio of the
standard deviation to the mean) over all benchmarks and all runs is 0.13, thus
the error bar is omitted from the visualization due to the little variance
compared to the performance difference shown.
Regardless of the metric, PROMPT reduces the profiling time by more than 90\%.
One source of the slowdowns in LAMP and Privateer is from building multiple
functionalities in one monolithic profiler.  This introduces unnecessary
functionality and the corresponding overhead.  PROMPT breaks down the profiling
tasks into modules each focusing on a single task. Note that targeting the
hottest loop, as PROMPT's profilers use, is another way of reducing the
unnecessary overhead.
PROMPT further optimizes the performance using parallelism in both the
address-based and the one built in the data structures like hash maps.

\paragraph{Improved Applicability to Complex Benchmarks.}

% \begin{figure*}[!ht]
%     \centering
%     \includegraphics[width=\textwidth]{figures/redesigned-spec2017.pdf}
%     \caption{The slowdown of the redesigned memory profiling modules on SPEC
%     2017 benchmarks.}
%     \label{fig:redesigned-comparison}
% \end{figure*}

As mentioned in \S\ref{sec:eval-speed}, the Privateer profiler fails or times
out on ten out of the 15 SPEC 2017 benchmarks. 
This is not a coincidence.  In fact, the clients using these memory profilers
are constrained by them, so they cannot evaluate on bigger benchmarks.  SCAF, a
system that shares the same memory profilers as Perspective, identifies that the
memory profilers are ``\textit{implemented in-house, lacking industrial-level
robustness in implementation}''~\cite{apostolakis:2020:SCAF}. Thus, it was
limited to only three SPEC 2017 benchmarks due to poor memory profiler
robustness or performance. 
Even for LAMP, which works for all SPEC 2017 benchmarks, or the Privateer, when
it works, the overhead is still significant as shown in \S\ref{sec:eval-speed}.
The robustness and the performance of the memory profilers are critical to the
applicability of memory profiling to more complex benchmarks.
The four memory profiling modules redesigned for Perspective exhibit greater
robustness and performance than their original counterparts. Three
modules (memory dependence, value pattern, and object lifetime) can run on all
SPEC 2017 benchmarks.
The points-to profiling module, which follows the same logic of
the parts of the original Privateer profiler which have design limitations,
fails on eight benchmarks. Two additional benchmarks work compared to Privateer
due to the memory allocation event hook in PROMPT which allow external calls
with memory allocation to be captured.
With the much-isolated codebase, we can also identify the root causes of
the failed benchmarks. The primary constraints include a lack of support for
\texttt{longjump/setjump} and the handling of non-null pointers to memory that
should not be dereferenced.  We are working on addressing these issues in a
future version of the module.

Performance-wise, the maximum slowdown for all modules is less than 35\Times{},
and most benchmarks are either below or around 10\Times{}, a huge improvement
over the original profilers discussed in \S\ref{sec:eval-speed}.
These overheads,
which translate to less than an hour of profiling time for benchmarks that
typically run for a few minutes, are sufficiently practical for users to test
clients using them. By enhancing the memory profiling workflow for more complex
benchmarks, PROMPT simplifies the adoption of systems like Perspective that rely
on memory profiling.

\subsection{Performance Analysis}
\label{sec:eval-perf-analysis}
% \begin{figure}[h]
%     \centering
%     \includegraphics[width=\columnwidth]{figures/overhead-reason.pdf}
%     \caption{The overhead of \tool as a memory dependence profiler with different configurations. The numbers are 
%     geomeans of all SPEC 2017 benchmarks.}
%     \label{fig:overhead-analysis}
% \end{figure}
\begin{table}[!ht]
    \small
    \centering
    % \resizebox{\columnwidth}{!}{%
    \begin{tabular}{ccc}
        \toprule
        \begin{tabular}[c]{@{}c@{}}PROMPT \\Optimizations\end{tabular} & \begin{tabular}[c]{@{}c@{}}Geomean \\ Slowdown\end{tabular} & Improvement \\ \midrule
        Baseline                      & 21.89\Times{}                 & N/A         \\ \midrule
        Specialization                & 14.48\Times{}                 & 51\%        \\ \midrule
        High-throughput Queue         & 12.29\Times{}                 & 18\%        \\ \midrule
        Data Parallelism              & 7.84\Times{}                  & 57\%        \\ \midrule
        High-throughput Data Structure & 7.26\Times{}                  & 8\%         \\
        \bottomrule
    \end{tabular}%
    % }
    \caption{Performance improvements with optimizations.}
    \label{tab:performance-optimization-comparison}
\end{table}

The performance improvement of PROMPT comes from designs discussed in 
\S\ref{sec:design} and \S\ref{sec:impl}.  In
Table~\ref{tab:performance-optimization-comparison}, we use the redesigned
memory dependence profiler in
Table~\ref{tab:PROMPT-and-perspective-loc-comparison} to show the effect of each
technique. 
All results are evaluated on SPEC 2017 benchmarks. 
The baseline is the memory dependence profiler without any optimization and we
incrementally apply each optimization to the baseline.  Note that these
improvement numbers are specific to this profiler.  Different memory profilers
may benefit differently from each technique.

\paragraph{Specialization}
\label{sec:eval-specialization}

\begin{table}[ht]
    \small
    \centering
    \begin{tabular}{ccccc}
        \toprule
        Profiler                       & \begin{tabular}[c]{@{}c@{}}Memory\\ Dependence \end{tabular} & \begin{tabular}[c]{@{}c@{}}Value \\Pattern\end{tabular} & \begin{tabular}[c]{@{}c@{}}Object \\Lifetime \end{tabular} & \begin{tabular}[c]{@{}c@{}}Points-\\to\end{tabular} \\\midrule
        % 500.perlbench & 44.3B & 21.6B & 25.1B & 5.0B & 14.0B \\\hline
        % 505.mcf & 96.9B & 56.1B & 51.2B & 19.8B & 39.5B \\\hline
        % 508.namd & 70.7B & 67.4B & 6.2B & 3.4B & 53.4B \\\hline
        % 510.parest & 329.7B & 168.1B & 198.3B & 116.2B & 120.2B \\\hline
        % 511.povray & 22.8B & 10.9B & 12.9B & 5.5B & 7.8B \\\hline
        % 519.lbm & 25.6B & 25.6B & 411.9M & 417.2M & 17.6B \\\hline
        % 520.omnetpp & 177.2B & 50.7B & 126.5B & 74.4B & 35.1B \\\hline
        % 523.xalancbmk & 280.3B & 93.4B & 196.2B & 117.6B & 64.5B \\\hline
        % 525.x264 & 185.1B & 134.8B & 54.3B & 85.1B & 80.6B \\\hline
        % 526.blender & 496.1B & 320.8B & 198.8B & 107.4B & 244.9B \\\hline
        % 531.deepsjeng & 109.4B & 79.2B & 40.1B & 34.8B & 51.4B \\\hline
        % 538.imagick & 114.1B & 113.0B & 10.6B & 9.9B & 66.1B \\\hline
        % 541.leela & 482.7B & 147.4B & 358.0B & 248.0B & 97.5B \\\hline
        % 544.nab & 9.0B & 6.5B & 2.2B & 1.5B & 5.0B \\\hline
        % 557.xz & 33.4B & 24.3B & 15.2B & 13.1B & 15.4B \\\hline
        \begin{tabular}[c]{@{}c@{}}Geomean Reduction (\%)\end{tabular} & 17.19                         & 54.04                         & 71.86                         & 52.89                         \\\bottomrule
    \end{tabular}
    \caption{The geomean reduction of profiling events with specialization for each memory profiler.
        % With specialization, the reduction of the profiling events is significant, ranging from 17\% to 72\%. 
    }
    \label{tab:specialization}
\end{table}

Table~\ref{tab:specialization} shows the reduction of the number of events with
specialization for different profilers.  With specialization, the reduction of
the profiling events is significant, ranging from 17\% to 72\%.

\paragraph{High-Throughput Queue}
\label{sec:eval-queue}
\begin{table}[!ht]
    % \small
    \centering
    \begin{tabular}{ccc}
        \toprule
        \multicolumn{2}{c}{Queue Type}                           & Time (ms)            \\ \midrule
        \multirow{2}{*}{boost::lockfree\cite{szuppe:2016:boost}} & queue       & 4603.7 \\
                                                                 & spsc\_queue & 555.1  \\ \midrule
        \multicolumn{2}{c}{Liberty Queue~\cite{jablin2010epic}}  & 48.6                 \\ \midrule
        \multirow{2}{*}{\textbf{PROMPT Queue}}                   & 1 Consumer  & 26.8   \\
                                                                 & 8 Consumers & 32.2   \\
        \bottomrule
    \end{tabular}
    \caption{The performance comparison of the queue.
        % evaluated with a benchmark that communicates ten million events.
        % The time is the average of 50 runs.
    }
    \label{tab:eval-queue-comparison}
\end{table}
We compare the performance of our queue implementation against others (two from
Boost~\cite{szuppe:2016:boost} and the Liberty queue~\cite{jablin2010epic}).  We
run it with a benchmark where two processes communicate ten million events from
the trace of \texttt{544.nab} through a shared-memory queue.  The
\texttt{boost::lockfree::spsc\_queue}, Liberty queue, and PROMPT queue are
configured with the same queue size (2MB).  The \texttt{boost::lockfree::queue}
is set to its max queue size of 65534.  We repeat the runs 50 times and take the
average.
As shown in Table~\ref{tab:eval-queue-comparison}, the PROMPT's queue
outperforms other queues by at least 81\%.  The performance improvement comes
from optimizations in \S\ref{sec:impl-queue} that reduce the overhead of event
production.  The throughput difference from one consumer to eight consumers is
only 20\%, a small cost to enable generic data parallelism.

\paragraph{Data Parallelism Wrapper}
\label{sec:eval-localwrite}

\begin{table}[ht]
    \small
    \centering
    \begin{tabular}{ccccccc}
        \toprule
        Parallel Workers            & 1    & 2    & 4   & 8   & 16  & 32  \\\midrule
        Geomean Slowdown (\Times{}) & 12.3 & 10.4 & 8.3 & 7.8 & 7.6 & 9.0 \\
        \bottomrule
    \end{tabular}
    \caption{The slowdown with different parallel workers with data parallelism wrapper for the memory dependence profiler.
    }
    \label{tab:data-parallelism}
\end{table}

Table~\ref{tab:data-parallelism} shows the slowdowns of the memory dependence
profiler with different numbers of workers for data parallelism.  The numbers
are the geomean slowdowns of all SPEC CPU 2017 benchmarks and the
high-throughput data structures are turned off for this evaluation. On the machine we
tested on, data parallelism improves the performance till 16 workers then starts
to drop.

\paragraph{High-throughput Data Structures}\label{sec:eval-container}

% We show the effect of the high-throughput container by evaluating it against a set of other container implementations.
%
We evaluate the performance with a benchmark that inserts ten million
dependences to \texttt{htmap\_count} that keeps the count of the
dependence~\S\ref{sec:impl-containers}.  The dependences are collected
from the trace of \texttt{544.nab}.  We run it ten times and take the average.
We compare the performance against two maps from C++ standard library, and
\texttt{phmap::flat\_hash\_map}, a more efficient open-source hash map
implementation from \texttt{parallel-hashmap} library based on
Abseil\cite{hashmap, abseil}.
The high-throughput map outperforms the standard library maps significantly; and
it outperforms \texttt{flat\_hash\_map} starting from two threads; with 32
threads, the performance almost doubles.
The time of the baseline shows if the insertion to the map is completely gone
and is the upper limit of our map.

\begin{table}[!ht]
    \small
    \centering
    \begin{tabular}{ccc}
        \toprule
        \multicolumn{2}{c}{Implementation}                  & Time (ms)                               \\ \midrule
        \multirow{2}{*}{libstdc++ (6.0.28)}             & \texttt{map}             & \textbf{319} \\
                                                        & \texttt{unordered\_map}  & \textbf{264} \\ \midrule
        Parallel Hashmap (1.3.8)\cite{hashmap}          & \texttt{flat\_hash\_map} & \textbf{102} \\ \midrule
        \multirow{6}{*}{\begin{tabular}[c]{@{}c@{}}PROMPT Data Structure\\ \texttt{htmap\_count}\end{tabular}} & 1                        & 126          \\
                                                        & 2                        & 91           \\
                                                        & 4                        & 75           \\
                                                        & 8                        & 70           \\
                                                        & 16                       & 60           \\
                                                        & 32                       & \textbf{53}  \\
                                                        & Baseline                 & 33           \\
        \bottomrule
    \end{tabular}
    \caption{The performance comparison of different implementations of maps to achieve a key to the count. The baseline of PROMPT \texttt{htmap\_count} only inserts to the buffer instead of inserting to the map.
        % evaluated with a benchmark that inserts ten million dependences into a map with the value as the count of the dependence.
    }
    \label{tab:container-comparison}
\end{table}

\paragraph{Memory and Binary Size Overhead}
\label{sec:memory-binary-overhead}
The memory overhead of the profiling frontend is the constant size oversize
introduced by the buffer of the queue.
The memory overhead of the backend comes from the constant size from the backend
code and data sections, the data structures to
store the profiling information, and auxiliary data structure during runtime.
Due to the reduction nature of the profiling process, the memory overhead of
the profiling information data structures is usually small.
The auxiliary ones depend on the implementation of the profiler.
A most significant and common cost comes from the shadow memory that
enables mapping from the address to the metadata.
The overhead of the shadow memory is bounded by \(P \times \text{heap memory
size} + \sum\text{profile size} + C\), where P is the shadow memory ratio
(number of bytes of metadata per byte of memory) and C is the constant cost
including the queue and other auxiliary data structures.
The data parallelism does not increase the memory overhead because the workers
share the same memory space.  We measured the peak memory overhead of the memory
dependence profiler running on all SPEC 2017 benchmarks.  When the fixed queue
is excluded, the backend memory overhead ranges from 20\% to 9.7\Times{}.
The instrumented binary size is 17\% to 231\% larger than the original.

%% file: sections/discussions.tex
\section{Discussion}
\paragraph{Potential Applications}
The most important application of PROMPT is speculative optimization.
While speculative optimizations, including automatic parallelization, have been
shown to be effective and broadly
applicable~\cite{apostolakis:2020:Perspective,bridges:07:micro,johnson:2012:Privateer,thies:07:micro},
these systems have not been widely adopted largely because of problems with memory profiling.
By reducing the runtime and engineering costs, PROMPT can greatly help
speculative optimization clients.
PROMPT also has the potential to attract a diverse range of users to build
various profilers on top of it or use it for different clients.  Multiple use
cases beyond speculative optimization can be addressed using the existing
profilers in PROMPT, such as memory prefetching, memory object layout
optimization, and security analysis.

\paragraph{Types of memory profiling not supported by PROMPT}
Memory profilers that alter the behavior of the program being profiled, such as
simulating the behavior of a hypothetical load instruction not present in the
original program (perhaps for prefetching), are not ideally suited for PROMPT.
While it is feasible to add new events to the frontend, as elaborated in
\S\ref{sec:impl-events}, it is crucial that these added events do not modify the
behavior of the program being profiled. They should only report such events, in
line with the design principles of PROMPT. We believe that most memory profiling
use cases can be addressed by solely implementing the profiling logic on the
backend, using existing profiling events.

\paragraph{Multi-threaded programs}
At present, PROMPT solely supports single-threaded programs, as its primary
motivation lies in speculative automatic parallelization clients that only
necessitate this level of support. To expand its capabilities for multi-threaded
workloads, events produced from multiple threads can either be combined into a
single queue or assigned to individual SPMC queues for each thread. The most
suitable approach depends on the requirements of the memory profiling modules. 

\paragraph{Profiling without source code available}
PROMPT provides full precision when the source code is available at compile
time. Functions from libraries that do not have source code available at compile
time are detected during compilation and reported to the client, who can then
decide how to proceed with the profiling results. In many cases, the profiling
results are still helpful but need to be conservative in cases involving
external calls.  A potential enhancement for PROMPT could involve incorporating
binary profiling. The decoupled design of PROMPT simplifies the implementation
process for such an addition.

\paragraph{Beyond memory profiling}
A memory profiler tracks memory-related events as listed in
Table~\ref{tab:profiling-events}.  Other types of profilers can be implemented with
this framework, as the list of possible events encompasses more than just memory
events. However, PROMPT's design is highly optimized for memory profiling. Other
profilers may not have as high a throughput as memory profiling and thus
may not benefit from PROMPT's queue and other optimizations.
The factorization process of memory profiling used in PROMPT, namely the separation
and generalization, may inspire other software systems. 
The separation helps to reduce the complexity, while the generalization helps to
reduce the cost of development. Both help with building a more efficient system.

% \subsubsection*{Sampling}
% Our motivation use case, speculative automatic parallelization, is very sensitive to precision of the memory profiler.
% Thus we build PROMPT to be fast without resorting to sampling.
% However, we do see value in sampling in some memory profilers that can be implemented with PROMPT.
% %
% With a decoupled design, PROMPT makes sampling easier to implement.
% %
% Sampling mechanism can be implemented as a part of frontend library with a boolean that controls whether
% the profile collection is on. 
% If the boolean is false, the profile collection is off, and events will not get generated to the queue.
% The logic of whether profile collection is on can be implemented anywhere as long as there is a way to alter the boolean.

% \subsubsection*{Multiple Profilers At the Same Time}
% The queue design also enables multiple memory profilers to consume from the same queue, however, we current do not implement the profilers in this way.

% \tool combines a targeting optimization with a shadow-memory based loop
% aware memory profiler.  This section describes the implementation of the
% clients, the targeting mechanic, the profiling run-time, as well as
% the parallelization of the profiling system.

%% file: sections/related.tex
\section{Related Work}
\label{sec:related}

% It is worth noting that some contributions that we make for a fast, robust and extendable LLVM-IR profiling framework are also valid for binary profiling.

%\subsubsection{LLVM Instrumentation}
%Other systems (Address/MemorySanitizer~\cite{asan, msan}, XRay~\cite{berris:2016:xray}, LLVM-Tracer~\cite{shao:2013:tracer}, Loom~\cite{loom:github}) implement instrumentation of LLVM IR.
% Other systems~\cite{asan, msan, berris:2016:xray,shao:2013:tracer,loom:github} implement instrumentation of LLVM IR.
% However, they are either too specific, focusing
% on a single task instead of generic memory profiling, or too generic, not able to use profiling oriented optimizations as in PROMPT.
% PROMPT is the first practical memory profiling framework that is both low-overhead and informative.
\paragraph{Memory Profilers}
%List the categories and use case
%List all memory dependence profiler
% Edge or hot profilers~\cite{gregg:2016:flame, preuss:2010:edgeprof} are used to identify hot functions and loops.
% These profilers are widely used for compiler optimizations and debugging. 
% PROMPT is a memory profiling framework that focuses on dynamic memory events.
% Both PROMPT and hot profilers may be used simultaneously to provide more information about the program runtime.

%Make it more smooth

Many memory profilers have been proposed for various use cases~\cite{apostolakis:2020:Perspective, kim:10:micro, yu:2012:fld, vanka:12:ead, mason2009lampview, johnson:2012:Privateer, zhao:2006:dep}.
They are different in terms of the profiling events they gather and the summarization method.
They can collect memory dependence~\cite{mason2009lampview,yu:2012:fld,kim:10:micro,zhao:2006:dep}, value pattern~\cite{gabbay:1997:valuepred}, object lifetime~\cite{wu:2004:obj}, and points-to relation~\cite{johnson:2012:Privateer}.
There are sub-variants for collecting memory dependence -- loop-aware, context-aware, tracking distance, or tracking counts\cite{chen:04:cc,CAMP,mason2009lampview,zhang:09:cgo}.
The growing number of different profilers also suggests new client profiling needs. 
PROMPT is an extensible memory profiling framework that can easily implement all these memory profilers.

Many directions have been explored to reduce the overhead of memory profiling.
Prior work has shown that shadow memory is particularly effective at improving run-time analysis of programs~\cite{zhao:10:ismm, nethercote:07:vee}.
Parallelism is also used in many memory profilers to optimize for speed~\cite{moseley:07:cgo,wallace:07:superpin, kim:10:micro,yu:2012:mcp}.
We leverage their findings and generalize their optimization in PROMPT.
% Yu et al.~\cite{yu:2012:fld} propose a technique that interleaves shadow
% memory with the actual application memory and also reduces the shadow memory size with type information. 
%
% CAMP~\cite{CAMP} statically builds the calling context tree in order to optimize the context manager. 
% While the current context manager of PROMPT is performant enough to service the Perspective client, this method could be used to further optimize PROMPT.
%
Lossy techniques can reduce overhead~\cite{vanka:12:ead, chen:04:cc}.
Vanka et al. combine sampling with a signature-based approach to achieve $3.0\times$ overhead~\cite{vanka:12:ead}.
% \cite{chen:04:cc} uses loop invocation sampling.
%to reduce the overhead of memory profiling.
However, such techniques suffer from imprecise results and some clients are very sensitive to precision.
PROMPT achieves low overhead without resorting to sampling.
Augmenting PROMPT with sampling for clients that tolerate imprecision is straightforward.

The LLVM address and memory sanitizer can be considered memory profilers with custom allocators~\cite{asan,msan}.
They achieve low overhead -- the address sanitizer reports less than 2.75x slowdown in the worst case and the memory sanitizer less than 7x.
However, their optimizations are very specialized for the given task and do not generalize to other memory profiling tasks considered in this paper.
PROMPT provides a framework on which various memory profilers can be built with generalized components and optimizations.

% Non-motivation
% There are also existing memory profilers for Python or garbage-collected languages like Java(CITE). The concept of memory profiling used in their settings is just to get the memory object statistics including max memory usage, which is a subset of the memory profiling we can do.

\paragraph{Implementing Memory Profilers}
%\subsubsection{Binary Instrumentation}
Pin and DynamoRio can instrument programs at the binary level~\cite{wallace:07:superpin,bruening:2012:dynamorio}.
LLVM and GCC have more freedom to instrument programs at the intermediate representation level~\cite{LLVM:CGO04,gcc}.
Tracing systems, sometimes built on top of instrumentation systems, collect program traces that can be used for online or offline analysis~\cite{zhang:04:micro, zhao:2006:dep, tallam:2007:UnifiedControlFlow, drcachesim}.
% However, binary instrumentation is not ideal for compiler optimizations. 
These systems help with building memory profilers.
However, even with these systems, building a memory profiler is hard. In addition, some tracing and binary instrumentation systems introduce baseline overheads for generating the trace or dynamic binary instrumentation.
PROMPT does not strive to replace instrumentation or tracing systems.
Instead, it focuses on memory profiling, providing components and optimizations to make building fast memory profilers much easier.
% PROMPT is an LLVM framework and provides much lower overhead.

% \cite{tallam:2007:UnifiedControlFlow} reports a 5x slowdown for collecting the trace, which is much faster than profilers that do analyses on the fly.
% PROMPT uses a design that is inspired by tracing systems where event collection and analyses are decoupled.
% This takes advantage of the relatively low slowdown of event collection and the fact that there is no feedback loop within the profiling framework, making it latency-insensitive.

\paragraph{Optimization Techniques}
% Odin~\cite{wang:2022:odin} is an LLVM instrumentation framework that allows flexible instrumentation. It does this by partitioning the program and recompiling the required code fragments as instrumentation needs change. This ensures that unnecessary probes are turned off. 
Program specialization to reduce cost has been proposed for many use cases~\cite{reps:1996:programSpecialization,schultz:2003:specialzationJava,wang:2022:odin}.
PROMPT uses a specialization technique where unnecessary events are not instrumented depending on the needs of the client.
PROMPT does it automatically at link time to remove the need to communicate with the LLVM pass.

The Liberty queue is the most related to the queue design~\cite{jablin2010epic, rangan2006pmup}.
It is a lock-free implementation designed for fast core-to-core communication and shifts communication overhead to the more idle end of the queue.
The PROMPT high-throughput queue design is influenced by the Liberty queue but leverages the latency-insensitive aspect of memory profiling to get more performance.
PROMPT uses a ping-pong buffer to reduce the cost of checking and communication and outperforms the Liberty queue by 81\%.
% The main difference is that memory profilers do not have a latency requirement, and the memory bandwidth of the main memory is enough for data transfer.
% The overhead comes from the additional cycles of the queue produce and consume instructions.
% \todo[inline]{Add additional queue implementation work}

Different techniques have been developed to make use of parallelism in memory profilers~\cite{moseley:07:cgo,wallace:07:superpin, kim:10:micro,yu:2012:mcp}.
PROMPT generalizes them as different forms of data parallelism and provides a generic data parallelism wrapper.
PROMPT automatically manages parallel workers and the interaction with shadow memory.
This makes it much easier to integrate data parallelism with any memory profiler.

The optimization used for the high throughput containers in PROMPT is parallel reduction~\cite{rauchwerger:95:sigplan}.
However, PROMPT wraps the parallelism in containers with insertion logic, so users can use them with ease and get parallelism for free.

%% file: sections/conclusion.tex
\section{Conclusion}
This paper presents a novel factorization of memory profiling, emphasizing the
significance of core profiling logic. This emphasis is achieved by separating the front and backend and generalizing shared
components and optimizations. Based on this factorization, the paper introduces
PROMPT, an open-sourced, fast, and extensible memory profiling framework.
Two existing LLVM-based memory profilers have been seamlessly ported to PROMPT,
resulting in simpler implementations and improved performance.
Furthermore, a tailored memory profiling workflow was redesigned for
Perspective, a state-of-the-art speculative parallelization framework. This
workflow is encapsulated in a concise 570 lines of code and reduces client
profiling time by more than 90\%. Such outcomes emphasize PROMPT's role in
enhancing the practicality and broader application of memory profiling
techniques.  In summary, this research positions PROMPT as a pivotal framework
in advancing the application of memory profiling techniques.